\documentstyle[aps,prl,multicol,epsf]{revtex}
\begin{document}

\title{
A new old algorithm, entropy sampling, and/or getting in and out of
energy minima}

\author{M.J.~Thill\\
Racah Institute of Physics\\
The Hebrew University of Jerusalem\\
91904 Jerusalem, Israel\\[2mm]
}

\maketitle


\begin{abstract}
I propose a new algorithm, a {\em free energy Monte Carlo algorithm}\/,  
for calculations where conventional Monte Carlo simulations 
struggle with ergodicity problems.
The simplest version of the proposed algorithm allows for the
determination of the entropy function of statistical systems and/or performs 
entropy sampling at sufficiently large times. I also mention how 
this algorithm can be used to explore the system's energy space,
in particular for minima.\\

{\noindent PACS numbers: 05.50.+q, 11.15.Ha, 64.60.Fr, 75.50.Hk,
87.10.+e, 02.60.Pn., 02.70.Lq, 05.20.-y\\[3mm]
}

\end{abstract}

\begin{multicols}{2}

\section{Introduction}

With fast-growing computer technology, Monte Carlo (MC) simulations have
with much success been used to study various statistical systems including
neural networks, problems in biology and chemistry, lattice-gauge
theories and optimisation problems in various areas, not to mention
statistical physics, to study the properties of phase transitions and critical
phenomena.

Most MC simulations concentrate on importance sampling for
the canonical or microcanonical Gibbs ensemble, introduced by Metropolis
{\em et al}\/ \cite{MRRTT53}. The thermodynamic average, $\overline{O}$, of an
observable $O(x)$ can be estimated \cite{F63} as
\begin{equation}\label{erg}
\overline{O}=\lim\limits_{t\to\infty}
\frac{\sum\limits_{\tau=t}^{t+t_a}O(x_\tau)P^{-1}(x_\tau)\exp(-\beta H(x_\tau))}
{\sum\limits_{\tau=1}^{t+t_a}P^{-1}(x_\tau)\exp(-\beta H(x_\tau))}\,\, ,
\end{equation}
where $x_\tau$ represents a configuration at time $\tau$ of a system with
Hamiltonian $H$, $\beta=(kT)^{-1}$ is the inverse temperature (with
Boltzmann constant $k$), $t_a$ an averaging time (with $t_a\gg 1$), and $P(x)$ a
sampling probability. If $P(x)$ is chosen to be constant, very few
samples contribute significantly to the sums in the above equation, and
a very long time is required to get a reasonable estimate of $\overline{O}$.
Importance sampling comes in if one chooses $P(x)$ as the Boltzmann
weight $\exp(-\beta H(x))$. It is generally a good sampling algorithm,
but can fail to access all the parts of the phase space in available
computer time. Indeed, in many situations,
this approach or similar ones face severe ergodicity problems if there
exist many high barriers between all the possible lowest- (or nearly-lowest-)
energy configurations, as, e.g., in certain Lennard-Jones systems
(see, e.g., \cite{VC72,TV77}) and in spin glasses (see, e.g.,
\cite{BY86,MPV,FH91,R95}), to cite only two examples from statistical physics.

To overcome these, at least for a large part, it has been suggested 
that it could be more efficient to reconstruct the Gibbs
ensemble from a simulation with other ensembles (e.g., a so-called
``multicanonical MC simulation'') than to simulate it
directly (see \cite{B95} and references therein). 
One of these approaches goes under the name of the {\em entropy sampling Monte Carlo
method}\/ (ESMC). It works as follows.
Let the probability of occurence of a configuration $x$ with energy $E$
be denoted as $P(x)$, and the probability of occurence of a state with
energy $E$ as $\tilde{P}(E)$. The term ``state'' stands here for the set of all
configurations that have the same energy. They are related to each other through
\begin{equation}
\begin{array}{lcrcl}
P(x) \propto e^{-\beta E}\,\, ,\\
\tilde{P}(E) \propto N(E) e^{-\beta E} = e^{S(E)/k-\beta E}\,\, ,
\end{array}
\end{equation}
where I have introduced the entropy, $S(E)$, of the state
with energy $E$. Their number is $N(E)$.
In the Metropolis MC method \cite{MRRTT53}, the canonical distribution of
states is obtained, along with ergodicity, by a Markovian sequence in
which the transition probabilities, $\pi(x\rightarrow x')$ and
$\pi(x'\rightarrow x)$, between a pair of configurations $x$
and $x'$ are determined by the {\em detailed balance}\/ condition
\begin{equation}
\frac{\pi(x\rightarrow x')}{\pi(x'\rightarrow x)}=\frac{\exp[-\beta
E(x')]}{\exp[-\beta E(x)]}\,\, .
\end{equation}
It can be shown rigorously that, in the case of a traditional MC 
simulation, this condition ensures a simulation of the system 
with an equilibrium distribution which is just the Gibbs
distribution \cite{vK81}. The ESMC method is, however, based on the
probability distribution of states, in which the probability of occurence
of a configuration with energy $E$ is proportional to the exponential of the 
{\em negative}\/ entropy, 
\begin{equation}
\begin{array}{lcr}
P(x) \propto e^{-S(E(x))/k}\,\, ,\\
\tilde{P}(E) \propto N(E) e^{-S(E)/k}\,\, .
\end{array}
\end{equation}
In a ESMC simulation the probability of occurence of a configuration
with energy $E$ is therefore anti-proportional to the number of
configurations with that energy. In this way, the probabilities of
occurence of all states equal the same constant. An MC algorithm that does the job
is one that is based on the detailed balance condition
\begin{equation}\label{dbs}
\frac{\pi(x\rightarrow x')}{\pi(x'\rightarrow
x)}=\frac{\exp[-S(E(x'))/k]}{\exp[-S(E(x))/k]}\,\, .
\end{equation}
In all other aspects, the formalism of the ESMC procedure follows 
then the usual Metropolis procedure. It is therewith easy to show (using
the methods exhibited, e.g., in \cite{vK81})
that the ESMC algorithm simulates the system in such a way that 
all states occur with the same probability. Hence, the 
algorithm provides for a (one-dimensional) {\em random walk 
through the system's energy space}\/. 

In this spirit, Monte Carlo sampling with respect to unconventional ensembles
has received some attention (see, e.g., \cite{B95} and references
therein) in recent years. In the ``multicanonical ensemble'' approach 
\cite{our1,our1a,our3}, one samples configurations such that the exact
reconstruction of canonical expectation values becomes feasible for a
desired temperature range. Multicanonical and related sampling 
has allowed considerable gains in situations with ``supercritical'' 
slowing down, such as (i) first order phase transitions
\cite{our1,our1a,our4} (for a recent review see, e.g., \cite{Janke}),
(ii) systems with conflicting constraints, such as spin glasses
\cite{our2a,temp,BHC,Ker1} or proteins \cite{HO,HS}.
The reconstruction of canonical expectation values requires 
knowledge of the entropy values of the/an important part of the
energy range (see equation (\ref{erg})), but leaves innovative 
freedom concerning the optimal shape \cite{Oxford}.
Considerable practical experience exists only for algorithms
where one samples such that (a) the probability density is flat 
in a desired energy range $P(E) = \mbox{const}$, and (b) 
each configuration of fixed energy $E$ appears with the same likelihood.
It should be noted that condition (b) is non--trivial. A simple algorithm
\cite{RC} exists to achieve (a), but which gives up (b). Exact connection
to the canonical ensemble is then lost. Such algorithms are interesting
particularly for hard optimisation problems, but may be 
unsuitable for canonical statistical physics. The present paper 
focuses on achieving (a) and (b).
To achieve a flat energy distribution, the appropriate unnormalised weight
factor in equation (\ref{erg}) is $S(E)$. However, 
before simulations, the entropy function $S(E)$ is usually
not known. Otherwise we would have solved the problem in the first place. Presumably,
reluctance about simulations with an a--priori unknown weight factor
is the main reason why the earlier umbrella sampling \cite{TV77} never
became popular in statistical physics.

In the more recent papers [9--30] 
it has been suggested to overcome this loophole by simulating 
with approximate entropy values, obtained by guessing
or a short Gibbs run, and then successively simulating with ever
better estimates of the real entropy. The underlying assumptions have, 
however, up to now never been shown to hold rigorously, nor is 
there anything known on convergence properties. 
In this paper I propose an algorithm from which the entropy can be obtained
in the large-time limit without attention ``by hand''. 
I have applied the algorithm to the
infinite-range, the two-dimensional and the three-dimensional
ferromagnet, spin glass, and for the search of the global minimum of
the generalisation error in some on-line learning model which exhibits
local minima besides a global one. In this letter, however, 
for illusatration and brevity, I concentrate mainly
on the infinite-range ferromagnet.
For definiteness, I will
formulate the algorithm for Ising spin systems, containing a total
number of $N$ spins. I will enumerate the different states of the 
system by their energy,
going from smallest to largest in value. Let $E(m)$ denote the
energy of the state with label $m$, say $m=0,1,\ldots,{\cal N}-1$ (so that
$\cal N$ is the total number of energy levels). Let furthermore $S(m;t)$ be the 
(estimated) entropy of the state with label $m$, at time $t$. 
At time $t=0$, I initialise $S(m;0)=0$ for all $m$. We will see below
that only entropy differences matter in the algorithm so that the
initialisation constant can in principle be chosen arbitrarily. The 
initialisation to zero is, however, preferable in the actual
implementation of the algorithm on a computer, as mentioned below.

The {\em Free Energy Monte Carlo}\/ (FEMC) algorithm then works as follows. 
Let us assume that, at time $t$, the system is in configuration $x$,
with energy $E(x)$, i.e., with label $m(x)$. 
Then, go through the following steps at time $t+1$:

\begin{enumerate}

\item Select one spin index $i$ for which the spin $s_i$ is considered
for flipping ($s_i\rightarrow -s_i$).

\item Calculate the transition probability 
\begin{equation}\label{tp}
\begin{array}{l}
\pi(x\rightarrow x';t+1)\\
{}\qquad:=\frac{1}{2}\left(1-\mbox{tanh}
\frac{S(m(x);t)-S(m(x');t)}{2}\right)
\end{array}
\end{equation}
to pass from configuration $x$ to configuration $x'$ which is obtained from $x$ 
by effectuating the considered spin flip. [In this present form the
algorithm is still rather an ``entropy Monte Carlo'' than a ``free
energy Monte Carlo'' algorithm; I will dwell on the full version of
the latter below.]

\item Draw a random number, $r$, uniformly distributed between zero and one.

\item If $r<\pi(x\rightarrow x';t+1)$, flip the spin, otherwise do not
flip it. In any case, the configuration of spins obtained at the end of
step 4.~is counted as the ``new configuration'', $x_{\mbox{\tiny update}}$. 

\item Now update the values of the entropy ($\mu=0,1,\ldots,{\cal N}-1$):
\begin{equation}\label{step5}
S(\mu;t+1) := S(\mu;t)+\epsilon(t) \delta_{\mu,m(x)}\,\, ,
\end{equation}
where $\epsilon(t)$ is a pre-chosen positive function which is
sufficiently small in the large-time limit, $m(x)$ the label of energy
of the configuration $x$, and $\delta$ denotes the Kronecker symbol. 

\item Go to 1. or end.

\end{enumerate}

Let us just note here that  
by the choice of $\pi(x\rightarrow x';t+1)$ as above, I ensured that
the simulation verifies a detailed balance condition ``locally'', i.e.,
at every time step. Furthermore, it can be proven \cite{t97} that
the algorithm converges indeed towards the entropy function of the
system in the infinite-time limit. Let us finally note that a
fter every time step, one should imagine,
using equation (\ref{step5}), that the ``zero'' (baseline of the entropy
values) has been shifted upwards. The entropy may then be obtained
as a time average over instantaneous entropy values, and normalising
these values with the help of the total number of configurations, $2^N$.

The above version of the algorithm does not deserve to be called
an FEMC algorithm just yet, as only the entropy enters. 
However, if one wants to detect the minima (or maxima)
in the energy space, it may be useful to change the transition probability 
of the algorithm to read:
\begin{equation}\label{FEMC}
\begin{array}{l}
\pi(x\rightarrow x';t+1)\\
\qquad:=\frac{1}{2}\left(1-\mbox{tanh}
\frac{S(m(x);t)-S(m(x');t)-\beta (E(x)-E(x'))}{2}\right)\,\, ,
\end{array}
\end{equation}
where now the transition probability does not depend solely on
(instantaneous) entropy differences, but on (instantaneous) {\em free
energy}\/ differences, $\beta$ being the ``inverse temperature'' as usual.
If $\beta$ is large, the temperature is small, and the system stays
preferably in configurations with low energy (if one were to take $\beta$
small or even negative (!) one stays obviously preferably in
configurations with high energy). This can be illustrated
in particular at the beginning of the algorithm, when the entropy 
differences are zero, and one performs a gradient descent algorithm
towards a local or global minimum out of which one is then taken
by a gradual increase in (instantaneous) entropy differences. 
In the following illustrations of the algorithm, 
I will, however, consider again the $\beta=0$
case only, for simplicity.

I have applied the algorithm to the infinite-range, the
two-dimensional and the three-dimensional ferromagnet to see 
how well the simplest version of the algorithm (the ``$\beta=0$ 
FEMC algorithm'') performs on obtaining the entropy values of the
considered systems and on overcoming energy barriers.
More specifically, in the case of the infinite-range ferromagnet, where
I know the entropy (or number of configurations) exactly,
I have investigated the convergence to the correct values
of the entropy as well as the passage times (``tunneling times'') 
between the ``all-spins up'' and ``all-spins down'' ground states.
The values of $\epsilon$ that I use are between $\epsilon=10^{-1}$
and $\epsilon=10^{-3}$. Certainly, $\epsilon$ should tend to zero
with increasing $t$ to obtain even more accurate values of the
entropy. However, the smaller $\epsilon$ is already in the
earlier stages of the algorithm, the longer it takes to
reach the asymptotic stage. If one takes $\epsilon$ too small at
the start of the algorithm, one risks to never leave, during the time of
the simulation, the regime where effectively one samples
according to performing a random walk in configuration space,
as the differences in the entropy values will stay too small
in the available simulation time. A good way to measure whether
one has reached the asymptotic regime of the algorithm yet or not
is to keep track of the sampled energies: if the histogram of the
energies, averaged over a long enough period of time, is flat,
asymptotics are reached. 

Some comments on the large-$\epsilon$ limit are also in place.
For $\epsilon$ of order $1$ or larger, the transition probability 
(\ref{tp}) of accepting a move becomes essentially $0$ or $1$, as the
differences in estimated entropy values for the different levels
become larger than $1$, hence the argument of the $\tanh$. This
has the following consequences. If, e.g., one is at time $t$, say, 
in a configuration of the level $\nu(t)$, and the values of the
estimated entropy values at the adjacent levels $\nu(t)\pm 1$
are $S(\nu(t)-1)<S(\nu(t))<S(\nu(t)+1)$, then moves are 
(essentially) accepted iff the level
of the considered update, $\nu_{\mbox{\tiny update}}$, is
$\nu_{\mbox{\tiny update}}=\nu(t)-1$. The algorithm can be 
viewed as putting a brick of height $\epsilon$ at every time 
step onto a wall which is building up during the
process. If the height of the wall at the 
adjacent place (level) whereto the move at time $t+1$ 
is considered is lower, the move is accepted with probability $\sim 1$,
if the wall is of equal height, the move is accepted with probability
$\frac{1}{2}$, and else it is (essentially always) rejected. 
It is easy to notice that on average the wall will be of equal
height for all levels if one substracts a running (increasing) baseline.
This leads to the observed fact that all energy levels occur
with equal probability, albeit the fact that the weight factors
in the algorithm are not proportional to the true entropy 
values (on average).

I have considered systems which contained $N=2^n$ spins where
$n=2,3,\ldots,9$. I have compared runs where the ``tunneling
times'' where measured from the beginning with ones where the
tunneling times of the first $x\cdot{\cal N}^2$ ($x=1,2,\ldots,10$)
Monte Carlo steps (MCS; defined as usual as the time needed to
update all $N$ spins in the system) were not taken into 
account. The (expected) experience from
these runs is that the distribution of tunneling times, its
mean ($\tau_{\mbox{\tiny m}}$) and random mean square (rms) value
($\tau_{\mbox{\tiny rms}}$) remained essentially 
unaltered. However, to be on the safe side, I have not
counted the tunnelings observed during the first $10\cdot{\cal N}^2$ MCS
in the runs whose results are displayed in the following.
In all of the different runs, the histogram of the
energies, averaged at the same time than the (instantaneous) entropy
values, is flat, i.e., fluctuates around the mean number of
sampled configurations per energy to within at most a small 
fraction of a percent for small system sizes and up to 
at most $\pm 5\%$ for the largest systems considered.

The results shown in figure~1 have been obtained with a total
number of $10^6+10\cdot{\cal N}^2$ MCS. I have fixed $\epsilon$
to equal $10^{-3}$, $10^{-2}$, and $10^{-1}$, respectively. This implies
that the values $\tau_{\mbox{\tiny m}}$ and $\tau_{\mbox{\tiny rms}}$,
the mean and the rms value of the tunneling times, have been obtained
for $\epsilon=10^{-3}$ from 149338 values for $n=4$ to 659 values for
$n=9$, for $\epsilon=10^{-2}$ from 148726 values for $n=4$ to 1218 
values for $n=9$, for $\epsilon=10^{-1}$ from 146682 values for 
$n=4$ to 1826 values for $n=9$. To check for statistical 
reliability of the data, I have also performed slightly 
shorter and longer runs. The error bars that I have got from 
these runs, at fixed averaging times (of the instantaneous entropy),
are smaller than the size of the symbols in the figures. 
The distributions for the tunneling times, $\tau$,
themselves are, however, intrinsically very broad 
(their width is of the order of their mean)
with an accordingly long tail, possibly power-law-like. 
In the runs displayed in the figures of this paper, at most $432$ 
tunneling events with times above $5\cdot\tau_{\mbox{\tiny m}}$ occured
for $N=2$, this number rapidly decreasing with system size to at most
$2$ such events for $N=512$. The statistics for
the larger systems may, indeed, be insufficiant because of the long tail
of the distribution (the effect of a reduction of the measured 
$\tau_{\mbox{\tiny m}}$ can clearly be seen in figure~1 for the
runs with $\epsilon=10^{-2}$ and $256$ respectively $512$ spins). 
However, the resulting error should not alter the results significantly.

Figure~1 shows the increase of $\tau_{\mbox{\tiny m}}$ with the
number of the spins in the system on a log--log scale. 
\begin{figure}
\begin{minipage}[t]{8cm}
\makebox[0cm]{}
        \epsfxsize=8cm	 
        \epsfbox{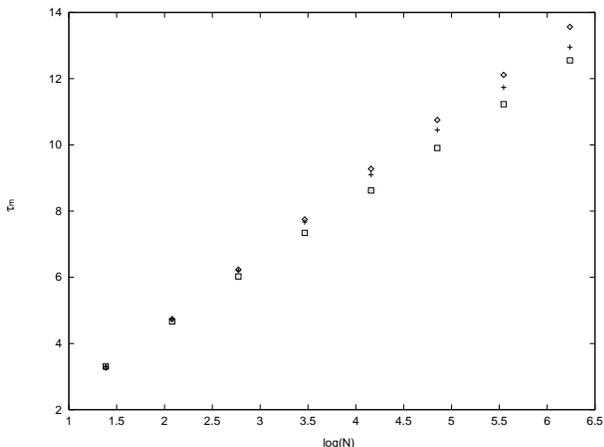}
\vspace{3mm}
\caption 
{Log-log-plot of the mean ``tunneling times'' 
$\tau_{\mbox \tiny m}$ versus the number of spins, $N=2^n$
($n=2,3,\ldots,9$), in the system for the infinite-range 
ferromagnet. The diamonds stem from simulations with $\epsilon=10^{-3}$,
the crosses from ones with $\epsilon=10^{-2}$, and the open squares
from ones with $\epsilon=10^{-1}$.}
\end{minipage}
\end{figure}
I have fitted straight lines (by eyesight) to the data points in figure~1, 
corresponding to the fits $\tau_{\mbox{\tiny m}}=c_{\mbox{\tiny m}} 
N^{\delta_{\mbox{\tiny m}}}$, for the different values of $\epsilon$. 
Because of higher statistical reliablitity of the results of the
smaller size systems these have been taken account more than the ones
of the larger size systems. The results for the fit parameters are
\begin{equation} 
\begin{array}{lcllcl}
\ln (c_{\mbox{\tiny m}}) \approx 0.36 \,\, , \quad  
\delta_{\mbox{\tiny m}} \approx 2.10 \,\, ,\qquad
\epsilon=10^{-3}\,\, ,\\
\ln (c_{\mbox{\tiny m}}) \approx 0.42\,\, ,  \quad \delta_{\mbox{\tiny m}} 
\approx 2.07 \,\, ,\qquad
\epsilon=10^{-2}\,\, ,\\
\ln (c_{\mbox{\tiny m}}) \approx 0.54\,\, , \quad \delta_{\mbox{\tiny m}} 
\approx 1.99\,\, ,\qquad \epsilon=10^{-1}\,\, .\\
\end{array}
\end{equation}
Remember that the exponent for a random walk in energy space 
is $\delta = 2$ from the general results on one-dimensional random
walks. In the above runs, we obtain that the width of the distribution 
is of the order of its mean; this can be noted easily already for
$N=1$, where one knows the (power-law-like) distribution of the tunneling
times and its properties analytically. 
 
In all of the above simulations, I have compared the obtained values
for the entropy with the true ones (that can be easily obtained exactly
[see, e.g., \cite{t97}]. Depending 
on the value of $\epsilon$, on the number of times that I averaged
over the (estimated instantaneous) entropy values and on the 
total time of the simulation, I got more and less accurate results.
In any case, for the runs of figure~1, the error was typically
of the order of $\epsilon$. In figure~1, a comparison of the entropy
values, obtained for a system containing $128$ spins, using the algorithm
with $\epsilon=10^{-2}$ and running it for a total of 50960 MCS, with 
the exact values is shown.
\begin{figure}
\begin{minipage}[t]{8cm}
\makebox[0cm]{}
        \epsfxsize=8cm	 
        \epsfbox{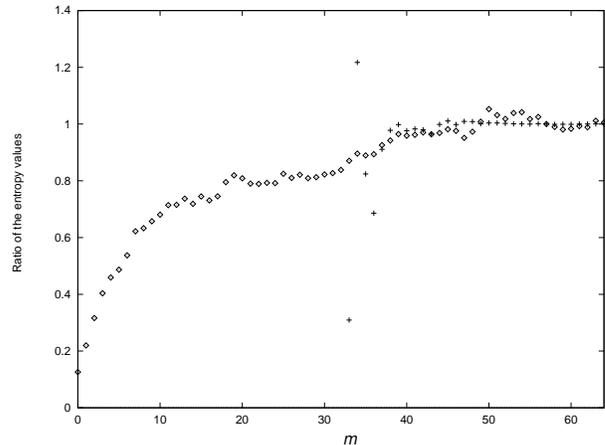}
\vspace{3mm}
\caption
{Comparison of the entropy values, $S(m)$, obtained through the FEMC algorithm
and $\epsilon=10^{-2}$ (crosses) with the true entropy values, and
comparison of the ones obtained through a random walk in 
configuration space (diamonds) with the true entropy values, 
for a system with 256 spins. The FEMC algorithm has been run in total for 
50960 MCS, the one of the random sampling for 1040960 MCS.}
\end{minipage}
\end{figure}
As can be seen from figure~2, the matching of the values (of the number
of configurations, not of the entropy~!) is quite
impressive, taking into account the fact that there are $2^{128}$
configurations. I have also compared it in the figure to the values 
obtained by a run with a larger number of MCS, namely 1040960 MCS,
but performing a local random walk in configuration space. In this 
last run, I have counted the configurations during the run and normalised
the sum of all the hits to $2^{128}$. It is easily
seen from the figure that, in contrast to the FEMC algorithm (with
much fewer MCS~!), ergodicity has been lost, configurations with 
energy $E(m)$ for $m<33$ have not even been sampled~!

In conclusion, in this paper, I have introduced a new algorithm in the spirit
of the proposals of simulations with ``multicanonical ensembles''.
I have simulated the infinite-range ferromagnet to obtain the entropy function 
and to investigate the ``tunneling times'' of getting from one
energy minimum to another one. The results are very encouraging: 
(i)The scaling of the ``tunneling times'' with the system size 
are roughly the ones of a random walk in energy space; 
(ii) the entropy function of the considered systems could be obtained
roughly within errors of order $\epsilon$ in the used computer time;
(iii) more importantly, ergodicity could be retained in all the
considered cases in the used computer time ($\sim 10^6$ MCS at most).
In particular, the last fact, the retaining of the ergodicity, should allow
for a calculation of physical quantities, such as correlation
functions, through equation (\ref{erg}) near zero temperature 
where conventional MC simulations fail. It is also interesting
to use the full FEMC algorithm (with nonvanishing $\beta$). More
results with the latter shall be published elsewhere \cite{t97}.
It is particularly interesting to determine the best choice of the
parameters $\epsilon$ and $\beta$ and of their ratio to overcome
energy barriers in the least amount of time and to explore the energy
space for minima or more generally extrema. Also, more 
analytical details and properties of the algorithm can be obtained 
for the case of the infinite-range ferromagnet \cite{t97}. It shall
be interesting to apply the algorithm in the cases where hitherto
ergodicity problems have not allowed to obtain results.

One last application of the algorithm should be mentioned here.
Constructing a general model of on-line learning 
is an important challenge in the theory of learning and its application. 
A plausible definition of the goal of supervised learning from examples is
to find a weight vector $\bf{w}$ that minimises the generalisation error, 
$\epsilon_g (\bf{w})$. In an on-line learning algorithm,
the learner receives a single new example at each time step 
and is unable to store previous examples in memory. 
The conventional on-line algorithm is based on the gradient of
the instantaneous error \cite{Hybook,hk-lpnn-91}.
For a sufficiently small learning rate, it converges
to a local minimum of  $\epsilon_g (\bf{w})$ but not necessarily to 
the global one. More importantly, it is not applicable to learning 
of boolean functions or of other discrete valued functions which are extremely 
useful for decision and classification tasks. 
Recently, an {\em on-line Gibbs algorithm}\/ has been proposed \cite{SK96}
as the first on-line algorithm that guarantees convergence 
to the minimal generalisation error for 
non-smooth systems, in particular for 
systems with discrete valued outputs or threshold hidden units. 
The price that is paid is an increased complexity of the computation at
each presentation of examples.
In particular, for systems in which the generalisation error has local minima, 
the on-line Gibbs algorithm may require a slow 
annealing schedule of the temperature variable, used in the algorithm,
which might yield a slow global convergence rate. 
In addition, the algorithm relies on the 
possibility of updating the weights by small increments. Consequently,
it is inapplicable to systems with {\em discrete valued weights}\/. 
In the case of the FEMC algorithm, gradient descent and escaping from 
minima can be combined naturally (see equation~(\ref{FEMC})) 
so that the FEMC may offer a valuable alternative to optimise the
learning procedure. Serious questions concerning 
on-line learning are still open, most importantly the one is on the 
{\em global convergence rate}\/, which is yet unknown in general.

\end{multicols}

\end{document}